\documentclass[10pt,pra,aps,floats,floatfix,twocolumn,showpacs,longbibliography,notitlepage,preprintnumbers,superscriptaddress]{revtex4-2}
\usepackage[USenglish]{babel}

\usepackage{amsmath,amssymb,amsfonts}

\usepackage{graphicx}% Include figure files
\usepackage[colorlinks=true,linkcolor=blue,citecolor=red]{hyperref}

\usepackage[T1]{fontenc} 
\usepackage{lmodern}

\usepackage[usenames,dvipsnames]{color}% color text

\newcommand{\sizetwo}{0.485\textwidth}
\newcommand{\sizethree}{0.45\textwidth}
\newcommand{\sizefour}{0.43\textwidth}

\begin{document}
\title{Ultrafast demagnetization in bulk nickel induced by X-ray photons\\ tuned to Ni $M_{3}$ and $L_3$ absorption edges} 
\author{Konrad J. Kapcia}
\email[e-mail: ]{konrad.kapcia@amu.edu.pl}
%\homepage[\mbox{ORCID ID}: ]{https://orcid.org/0000-0001-8842-1886}
\homepage[\mbox{ORCID ID}: ]{0000-0001-8842-1886}
\affiliation{\mbox{Institute of Spintronics and Quantum Information, Faculty of Physics}, Adam Mickiewicz University in Pozna\'n, 
Uniwersytetu Pozna\'{n}skiego 2, 61614 Pozna\'{n}, Poland}
\affiliation{\mbox{Center for Free-Electron Laser Science CFEL, Deutsches Elektronen-Synchrotron DESY}, Notkestr. 85, 22607 Hamburg, Germany}

\author{Victor Tkachenko}
\email[e-mail: ]{victor.tkachenko@xfel.eu}
\homepage[\mbox{ORCID ID}: ]{0000-0002-0245-145X}
\affiliation{European XFEL GmbH, Holzkoppel 4, 22869 Schenefeld, Germany}
\affiliation{\mbox{Center for Free-Electron Laser Science CFEL, Deutsches Elektronen-Synchrotron DESY}, Notkestr. 85, 22607 Hamburg, Germany}

\author{Flavio Capotondi}
\affiliation{Elettra-Sincrotrone Trieste S.C.p.A, 34149 Trieste, Basovizza, Italy}

\author{\mbox{Alexander Lichtenstein}}
\affiliation{European XFEL GmbH, Holzkoppel 4, 22869 Schenefeld, Germany}
\affiliation{University of Hamburg, Jungiusstr.  9, 20355 Hamburg, Germany}

\author{\mbox{Serguei Molodtsov}}
\affiliation{European XFEL GmbH, Holzkoppel 4, 22869 Schenefeld, Germany}
\affiliation{Institute of Experimental Physics, TU Bergakademie Freiberg, Leipziger Strasse 23, 09599 Freiberg, Germany}
\affiliation{Center for Efficient High Temperature Processes and Materials Conversion (ZeHS), TU Bergakademie Freiberg, Winklerstrasse 5, 09599 Freiberg, Germany}

\author{Przemys\l{}aw Piekarz}
\affiliation{\mbox{Institute of Nuclear Physics, Polish Academy of Sciences},  Radzikowskiego 152, 31-342  Krak\'ow, Poland}

\author{Beata Ziaja}
\email[e-mail: ]{ziaja@mail.desy.de}
\homepage[\mbox{ORCID ID}: ]{0000-0003-0172-0731}
\affiliation{\mbox{Center for Free-Electron Laser Science CFEL, Deutsches Elektronen-Synchrotron DESY}, Notkestr. 85, 22607 Hamburg, Germany}
\affiliation{\mbox{Institute of Nuclear Physics, Polish Academy of Sciences},  Radzikowskiego 152, 31-342  Krak\'ow, Poland}
\date{\today}
\begin{abstract}
Studies of light-induced demagnetization started with the experiment performed by Beaupaire et al.  on nickel. Here, we present theoretical predictions for X-ray induced demagnetization of nickel, with X-ray photon energies tuned to its $M_3$ and $L_3$ absorption edges. We show that the specific feature in the density of states of the d-band of Ni, a sharp peak located just above the Fermi level, strongly influences the change of the predicted magnetic signal, making it stronger than in the previously studied case of cobalt. We believe that this finding will inspire future experiments investigating magnetic processes in X-ray irradiated nickel. 
\end{abstract}

%\pacs{}

\maketitle

%========================================================================
%%%%%%%%%%%
\section{Introduction}

Ultrafast control of magnetization with lasers remains a hot topic in laser and solid-state physics communities. Apart from traditional terahertz and optical lasers, the state-of-art XUV or X-ray free-electron lasers \cite{Acker2007,Alla2012,Emma2010,Pile2011,Weise2017} are now also used for demagnetization studies. The main advantage of these lasers is possibility to resonantly excite core electrons to the magnetically sensitive d-band. As the electronic occupation in the the d-band determines the magnetization of the material, the X-ray induced electronic excitation changes the population of spin-up and spin-down electrons in the band. This results in the decrease of the total magnetic moment in the material \cite{KapciaNPJ2022,KapciaPRB2023,KapciaCCC2023}. In our previous studies \cite{KapciaNPJ2022,KapciaPRB2023}, we modeled the experimentally observed ultrafast decrease of the X-ray scattering signal from X-ray irradiated cobalt which reflected a transient decrease of the cobalt magnetic moment. The XSPIN simulation tool was developed to follow the progressing demagnetization of cobalt.  Our studies have shown that the signal decrease can be explained by ultrafast electron-driven demagnetization. 

In this paper, we will apply our model to another widely-used magnetic material, nickel. Magnetic moments of nickel and cobalt are $0.66\, \mu_B$ and $1.70\, \mu_B$ respectively \cite{MeyerJPC2015}. As nickel's Curie temperature ($627$ K) \cite{Chatt1978} strongly differs strongly from that of cobalt ($1400$ K), such study can reveal a potential effect of the Curie temperature on the demagnetization dynamics. Laser triggered demagnetization of nickel has been studied in various papers, see, e.g., \cite{BeaurepairePRL1996,KoopmansPRL2000,StammNatMat2007,LojewskiMRL2023,KriegerJCTC}. Interestingly, so far, we have not found any relevant experimental data on Ni demagnetization recorded at XFEL facilities. Therefore, the actual comparison between Co and Ni demagnetization will be performed with theoretical predictions only.

%%%%%%%%%%%%%%%%
\subsection{Simulation scheme}

As in our previous works \cite{KapciaNPJ2022,KapciaPRB2023,KapciaCCC2023}, we will use our recently developed XSPIN code to obtain predictions for the 'magnetic signal' from the X-ray irradiated nickel. The electronic density of states is obtained from the density functional theory  (DFT) calculations implemented in the  Vienna Ab initio Simulation Package (VASP) \cite{VASP1,VASP2,VASP3}. Average absorbed doses considered in the simulations are chosen not to cause structural changes (atomic dislocations) in the irradiated materials. Therefore, the equilibrium density of states (DOS) can be used throughout the whole simulation (the ''frozen atom'' assumption). The occupations of electronic levels change during the material exposure to the X-ray pulse, as due to the photoionization, impact ionization and Auger process, excited electrons leave the band to the continuum. Later,  they relax back to the band. As the electrons are heated up by the pulse, they remain hot on femtosecond timescales considered in this study, as their temperature can only decrease through an exchange with the lattice which follows on longer ((sub)ps) timescales. Moreover, due to the assumed common thermalization of all electrons, both spin-up and spin-down ones (following Fermi Dirac distribution with a common temperature and a chemical potential), the numbers of spin-up and spin-down electrons will be different from the corresponding values in the initial state. This thermalization-induced spin flip process, changing the population of spin-up and spin-down electrons, leads to a change of the magnetic signal.

For the simulation, we use a simulation box with $N = 512$ Ni atoms. We provide averaging over $100\ 000$ realizations in the Monte Carlo module. The XFEL pulse is assumed  to have a Gaussian temporal profile of the duration of $70$ fs FWHM (full width at half maximum) for M-edge case (M$_3 = 66.2$ eV) and $50$ fs FWHM for L-edge case (L$_3 = 852.7$ eV). The pulse duration was chosen such to compare the XSPIN predictions for nickel with our previous results for cobalt presented in \cite{KapciaNPJ2022,KapciaPRB2023,KapciaCCC2023}. For more details on the simulation parameters, see Tab. \ref{tab:one}.
%%%%%%%%%%%%%%%%%%%%% 

%%%%%%%%%%%%%%%%%%%%%%%%%%%%%%%
\begin{table*}[!t]
\caption{\label{tab:one} Simulation parameters used in the present work. All energies are in electronovolts (eV).}
%\begin{ruledtabular}
\begin{tabular}{ccccc }
Region       $\ $ & $\ $ Photon energy $\hbar \omega_\gamma$ $\ $ & $\ $ Probed level $\hbar \omega_0$ $\ $ & $\ $ $\Delta$ $\ $ & $\ $ Energy range $ [\hbar\omega_0 - \Delta ; \hbar\omega_0 + \Delta] $ \\
\hline
M-edge &    $67$  &   $0.8$     &   $0.7$ & $[-0.1; 1.5]$ \\
M-edge &    $68$  &   $1.8$     &   $0.7$ & $[ 1.1; 2.5]$ \\
L-edge &    $853$  &   $0.3$     &   $1.0$ & $[-0.7; 1.3]$ \\
L-edge &    $854$  &   $1.3$     &   $1.0$ & $[ 0.3; 2.3]$ 
\end{tabular}
%\end{ruledtabular}
\end{table*}
%%%%%%%%%%%%%%%%%%%%%%%%%%%%%%%

%%%%%%%%%
\section{Results}

%%%%%%%%%%%%%%
\subsection{Spin-polarized electronic density of states from density functional theory calculations}

In order to obtain spin-polarized electronic density of states for bulk nickel, we performed  first-principle calculation, using the projector augmented wave (PAW) potentials \cite{BlochPRB1994} and the generalized gradient approximation (GGA) in the Pardew, Burke, and Ernzerhof (PBE) parametrization \cite{PerdewPRL1997}, implemented in the VASP code \cite{VASP1,VASP2,VASP3}.

\begin{figure}
\includegraphics[width=\sizetwo]{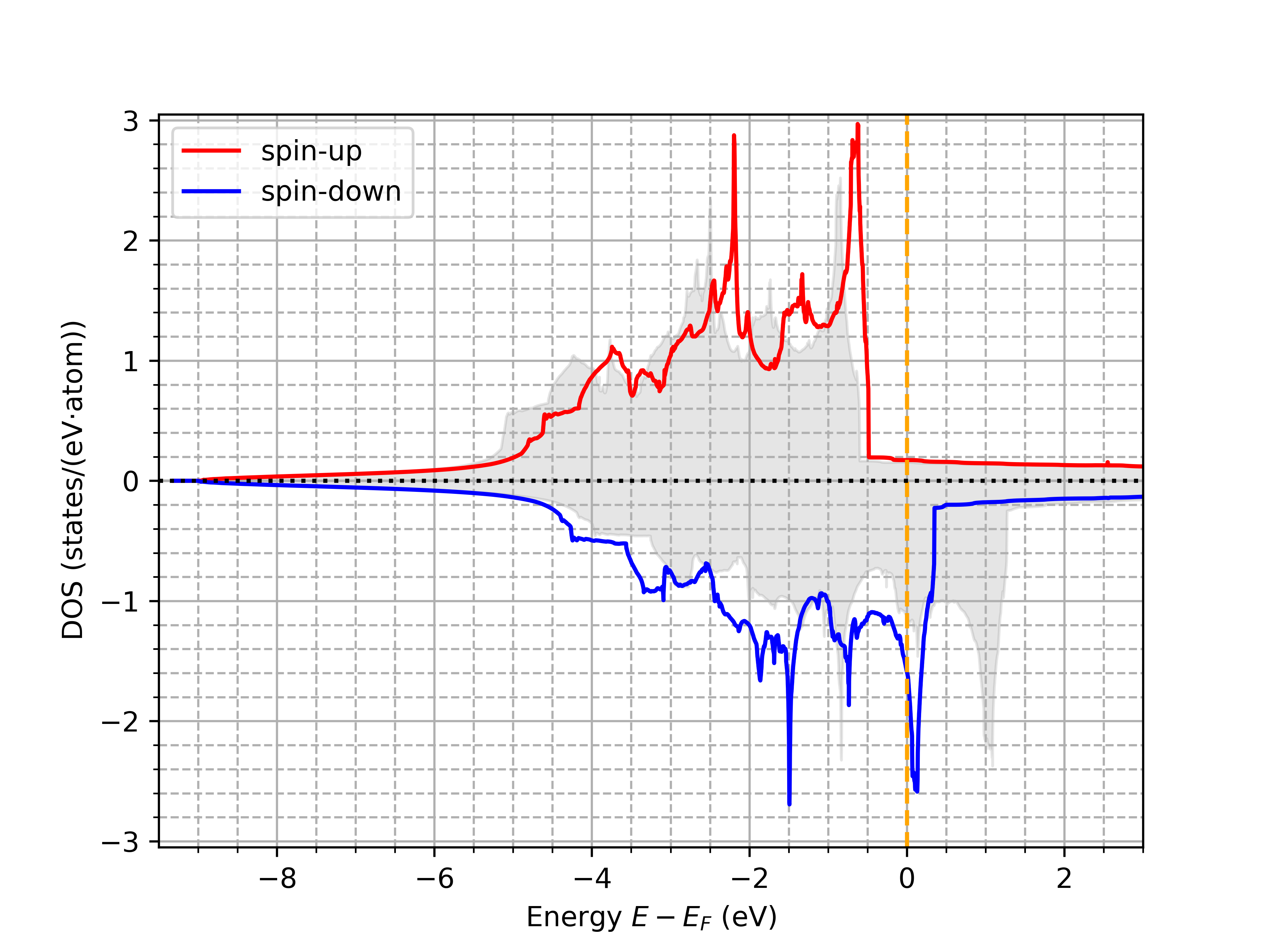}
\caption{\label{fig:DOS}%
Calculated density of states (states per eV, per atom) for fcc nickel. The red and blue lines correspond to DOS for spin-up and spin-down electrons. The grey shadowed area indicates the density of states for fcc cobalt investigated in Refs. \cite{KapciaNPJ2022,KapciaPRB2023,KapciaCCC2023}.
}
\end{figure}

For the summation over the reciprocal space, we used  $27 \times 27 \times 27$ Monkhorst-Pack $k$--point grid \cite{MonkhorstPRB1976}. The spin-polarized density of states for fcc bulk Ni (calculated for the experimental bulk value of the lattice constant, $a = 3.524\ \text{\AA}$) is presented in Figure \ref{fig:DOS}. It is in an agreement with other DFT calculations (see also, e.g., \cite{AbdallahAIP2014}). For comparison, the density of states for fcc bulk Co (with $a = 3.545\ \text{\AA}$ \cite{Wang2019}) used in Refs. \cite{KapciaNPJ2022,KapciaPRB2023} is also presented.
The calculated magnetic moments of nickel and cobalt are $0.62\, \mu_B$ and $1.61\, \mu_B$, respectively i.e., with a good agreement with those from \cite{MeyerJPC2015}.

%%%%%%%%%%%%%%%%%%%%%%%%%%%%%%%
%%%%%%%%%%%%%%%%%%%%%%%%%%%%%%%

%%%%%%%%%%%%%%%%
\subsection{Electronic properties of X-ray irradiated nickel}

\begin{figure*}
(a) \includegraphics[width=\sizethree]{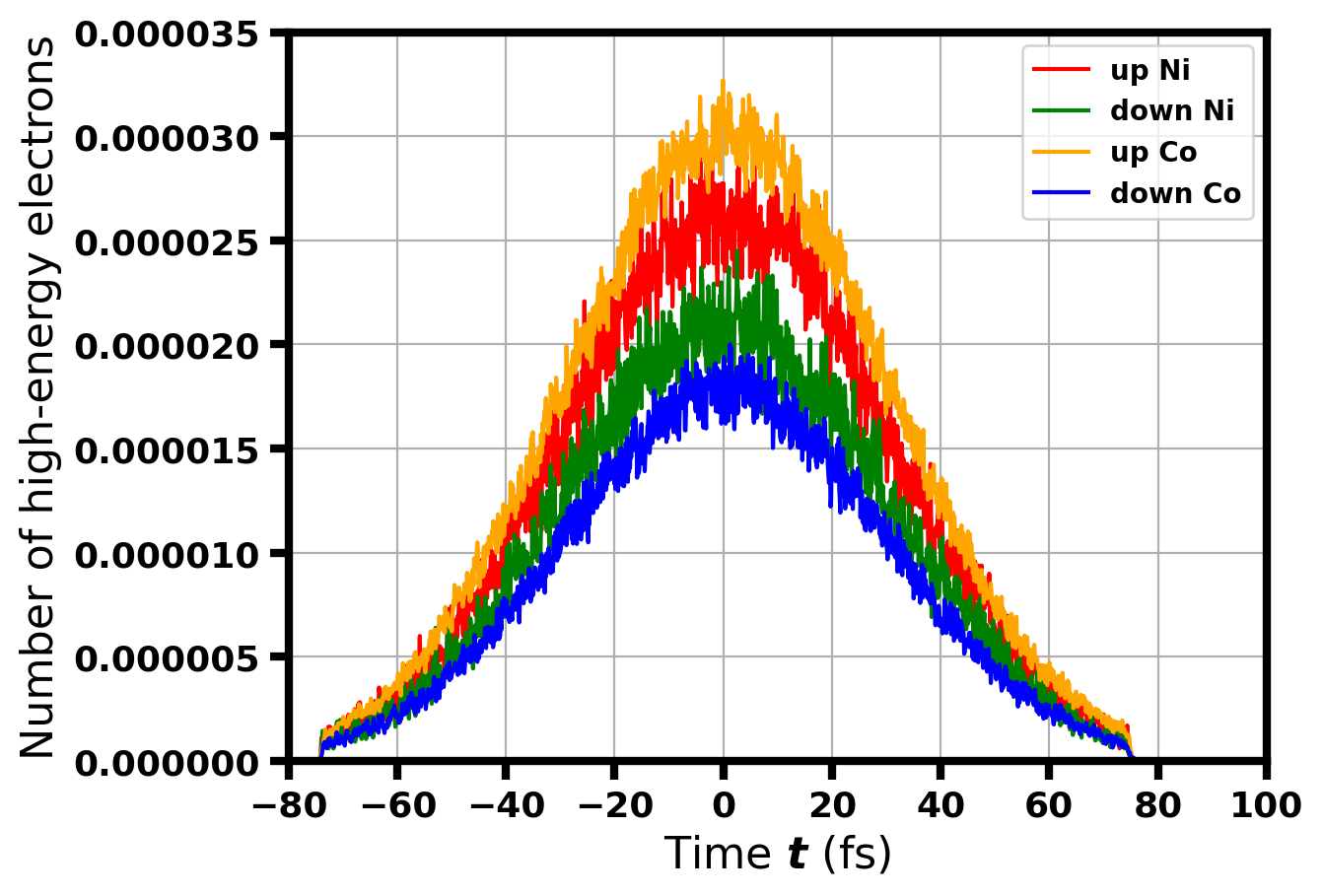}
(b) \includegraphics[width=\sizethree]{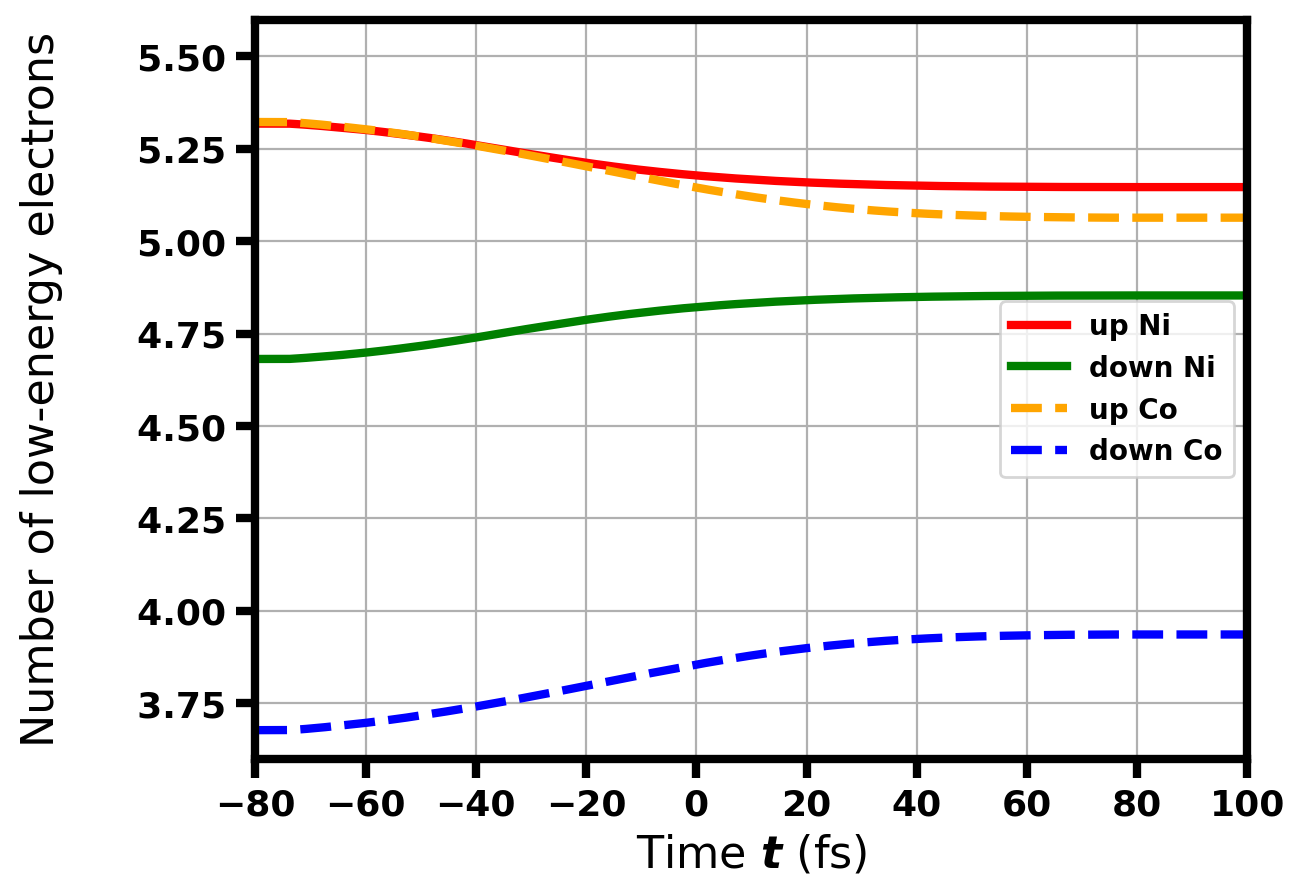}\\
(c) \includegraphics[width=\sizethree]{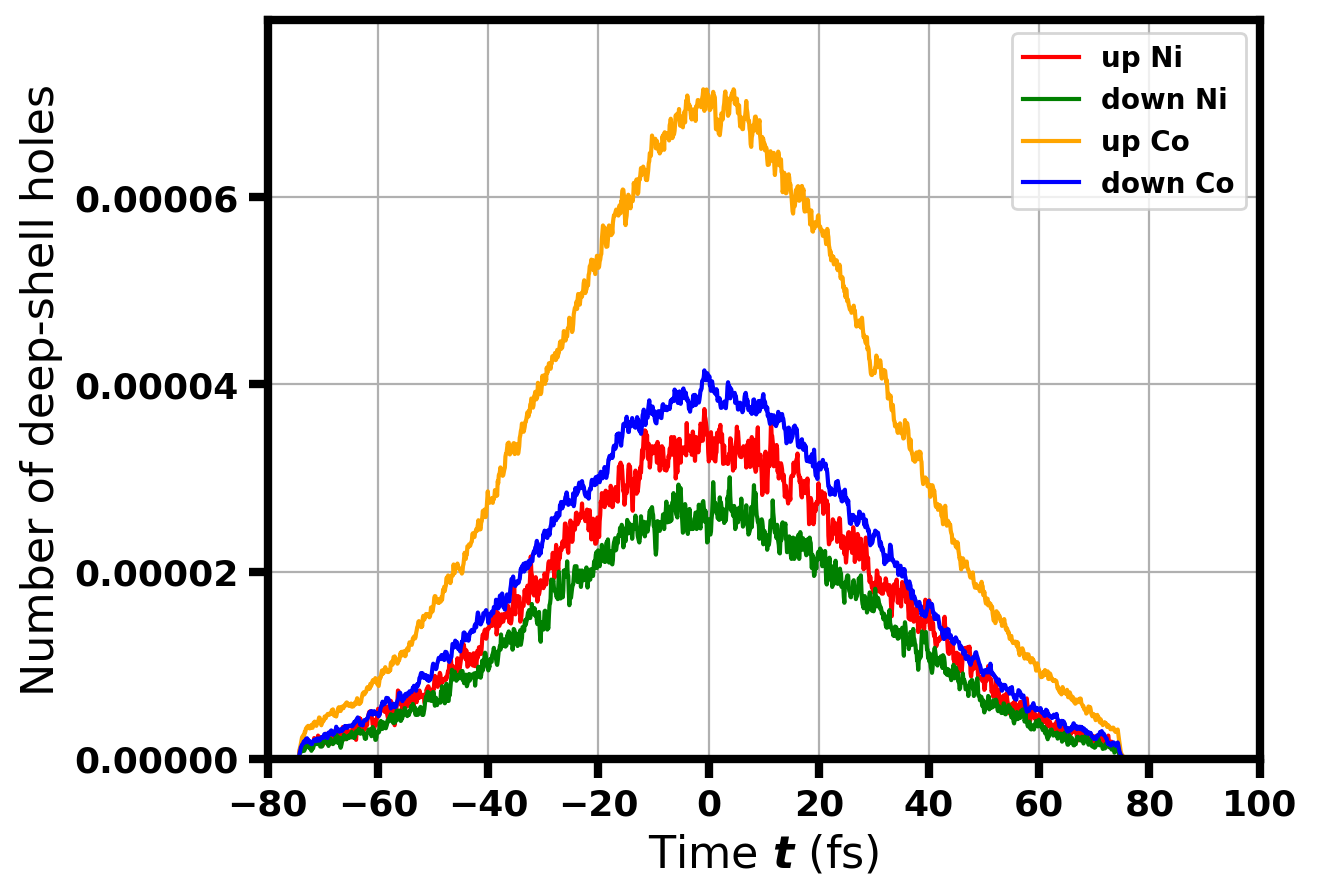}
(d) \includegraphics[width=\sizethree]{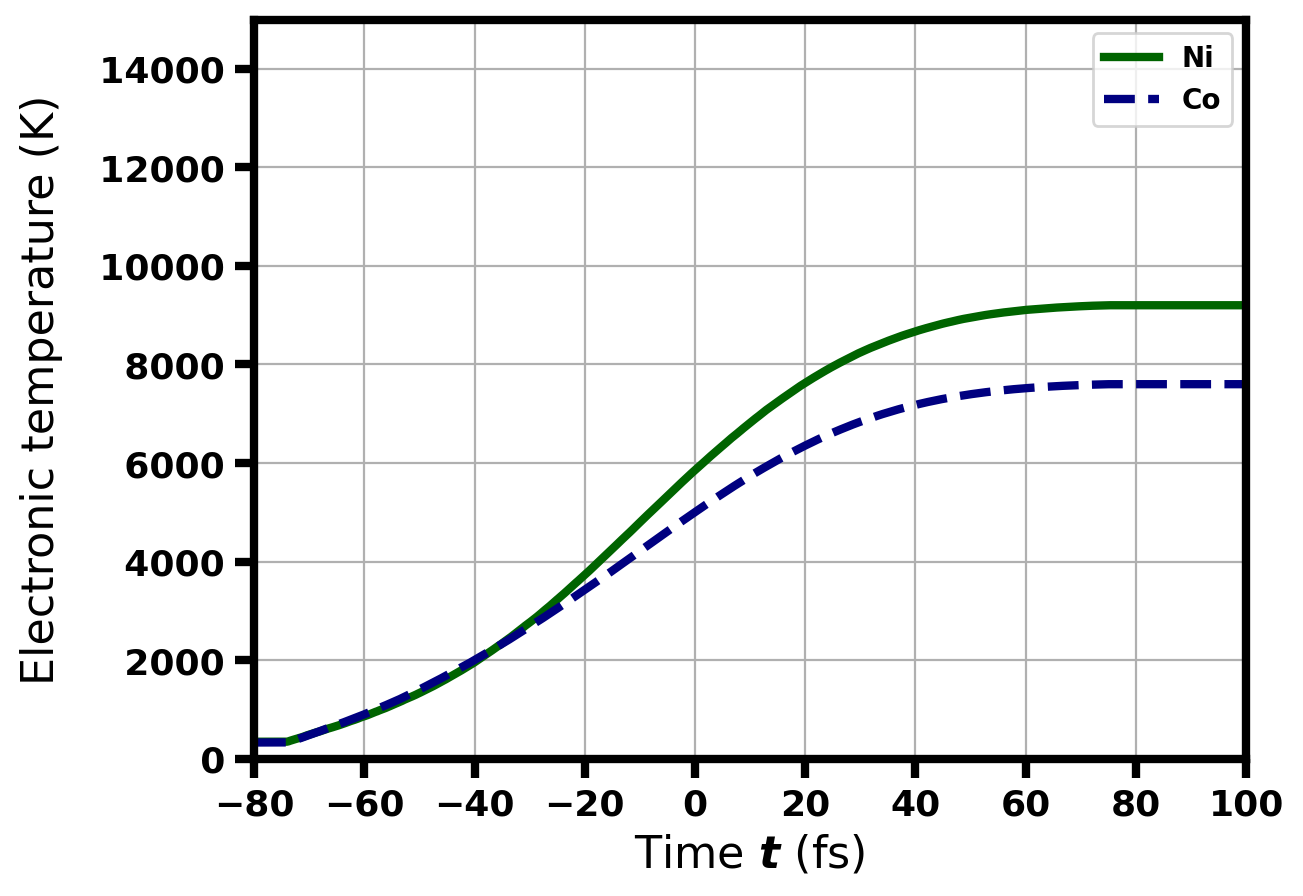}
\caption{\label{fig:NivsCo}
Transient distributions of excited electrons and holes obtained with XSPIN code for X-ray irradiated nickel and for cobalt:  (a) the transient number of polarized high energy electrons (with energies $> 15$ eV),  (b) the number of low energy electrons (with energies $< 15$ eV), (c) the transient number of deep shell holes (with indicated polarization of electrons previously occupying the holes), and (d) electronic temperature. The X-ray photon energy was tuned to M-edge of Ni (67 eV) and to M-edge of Co (61.1 eV). Pulse duration was 70 fs FWHM. Average absorbed dose was 0.93 eV/atom.}
\end{figure*}

Below we present the results on the transient distributions of  excited electrons and holes obtained with the XSPIN code for nickel and for cobalt (cf. also \cite{KapciaNPJ2022,KapciaPRB2023}) irradiated with X rays tuned to their M absorption edges ($\sim 67$ eV and $\sim 61$ eV respectively).  Figure \ref{fig:NivsCo} shows: (a) the transient number of polarized high energy electrons (with energies $> 15$ eV),  (b) the number of low energy electrons (with energies $< 15$ eV), (c) the transient number of deep shell holes (with indicated polarization of electrons previously occupying the holes), and (d) electronic temperature. The photoexcitation dynamics in Co and Ni look qualitatively similar, with a stronger excitation in Co (Figure \ref{fig:NivsCo}a-b) than in Ni. Collisional relaxation in Ni is also weaker than in Co (Figure \ref{fig:NivsCo}c), which leads to the higher electronic temperature in Ni, when compared to Co (Figure \ref{fig:NivsCo}d). 

%%%%%%%%%%%%%%%%%%%%%%
%%%%%%%%%%%%%%%%%%%%%%
%%%%%%%%%%%%%%%%%%%%%%%

\subsection{Generalized transient magnetization}

In order to follow changing magnetic properties of irradiated materials, we have introduced in \cite{KapciaNPJ2022} a generalized transient magnetization which reflects the disparity between electronic populations in spin-up and spin-down electronic subsystems in the d-band:
%%%%%%%%%%%%%%%%%%%
\begin{equation}\label{eq:genmagnet}
M(t) = \sum_{\hbar \omega_0 - \Delta}^{\hbar \omega_0 + \Delta}  \left[ N^{\textrm{h}}_{\uparrow} (E_{i,\uparrow}) - N^{\textrm{h}}_{\downarrow} (E_{i,\downarrow})\right],
\end{equation}
%%%%%%%%%%%%%%%%%%%
The probed region in $d$-band extends between $\hbar \omega_0 - \Delta$ and $\hbar \omega_0 + \Delta$, where $\hbar \omega_0 = \hbar \omega_\gamma - E_{\textrm{edge}}$. Here, $\hbar \omega_\gamma$ is the incoming photon energy, and $E_{\textrm{edge}}$ is the energy of the resonant core $p$-level. The summation goes here over discrete levels. Note that we neglect the subleading effect of the different coupling of polarized light to spin-up and spin-down electrons (XMCD)  here. 
Electronic populations are calculated, assuming Fermi-Dirac distribution of electrons. Knowing at every time step $t$ electronic temperature $T_{\textrm{e}}$ and electronic chemical potential $\mu$,  we have $ N^{\textrm{h}}_{\sigma} (E) = 1 - N_{\textrm{e},\sigma}^{\textrm{low}}(E)$ and $N_{\textrm{e},\sigma}^{\textrm{low}}(E) = \left\{ 1 + \exp \left[ (E-\mu)/k_{\textrm{B}} T_{\textrm{e}} \right] \right\}^{(-1)}$.

\begin{figure*}
\includegraphics[width=\sizefour]{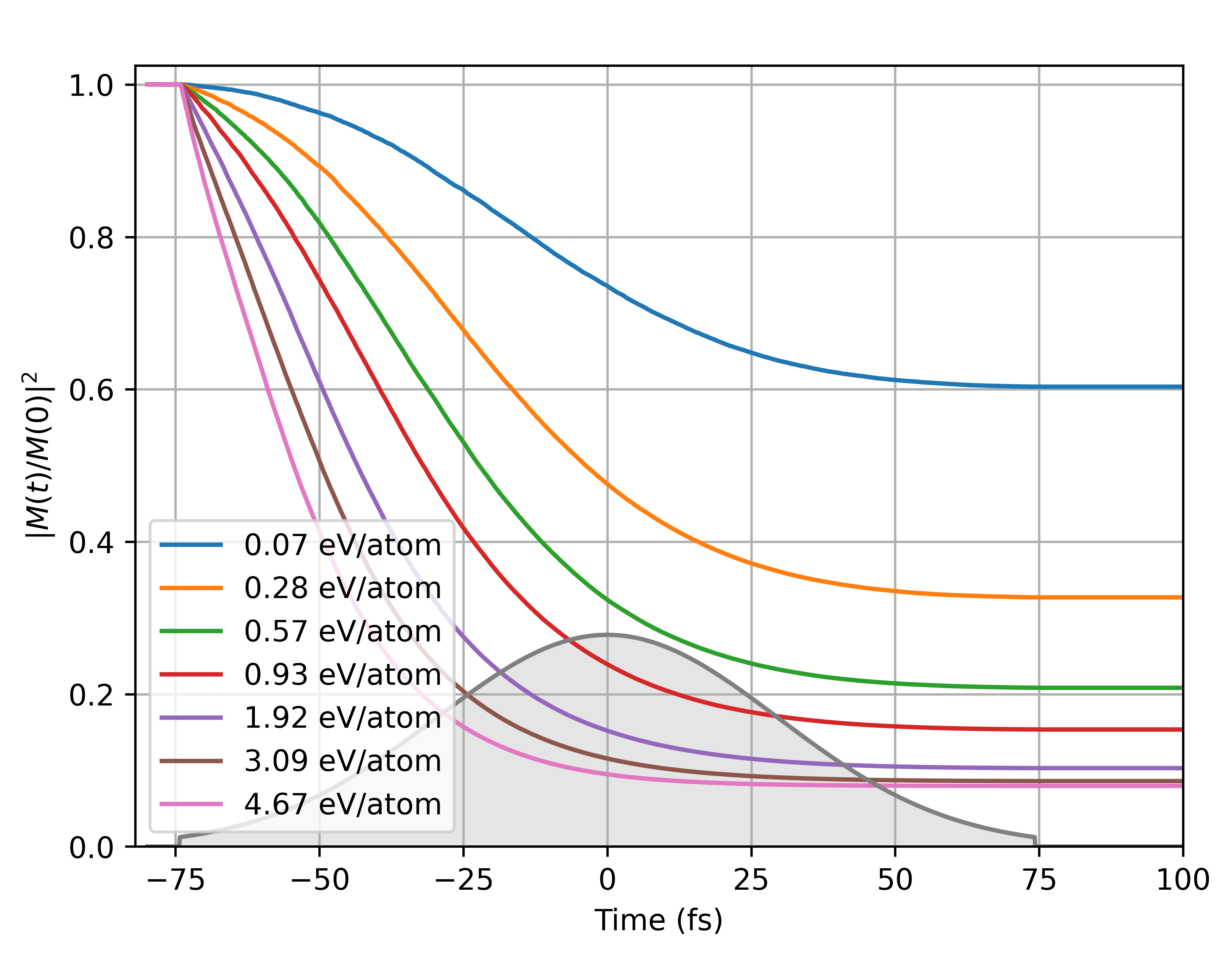}
\includegraphics[width=\sizefour]{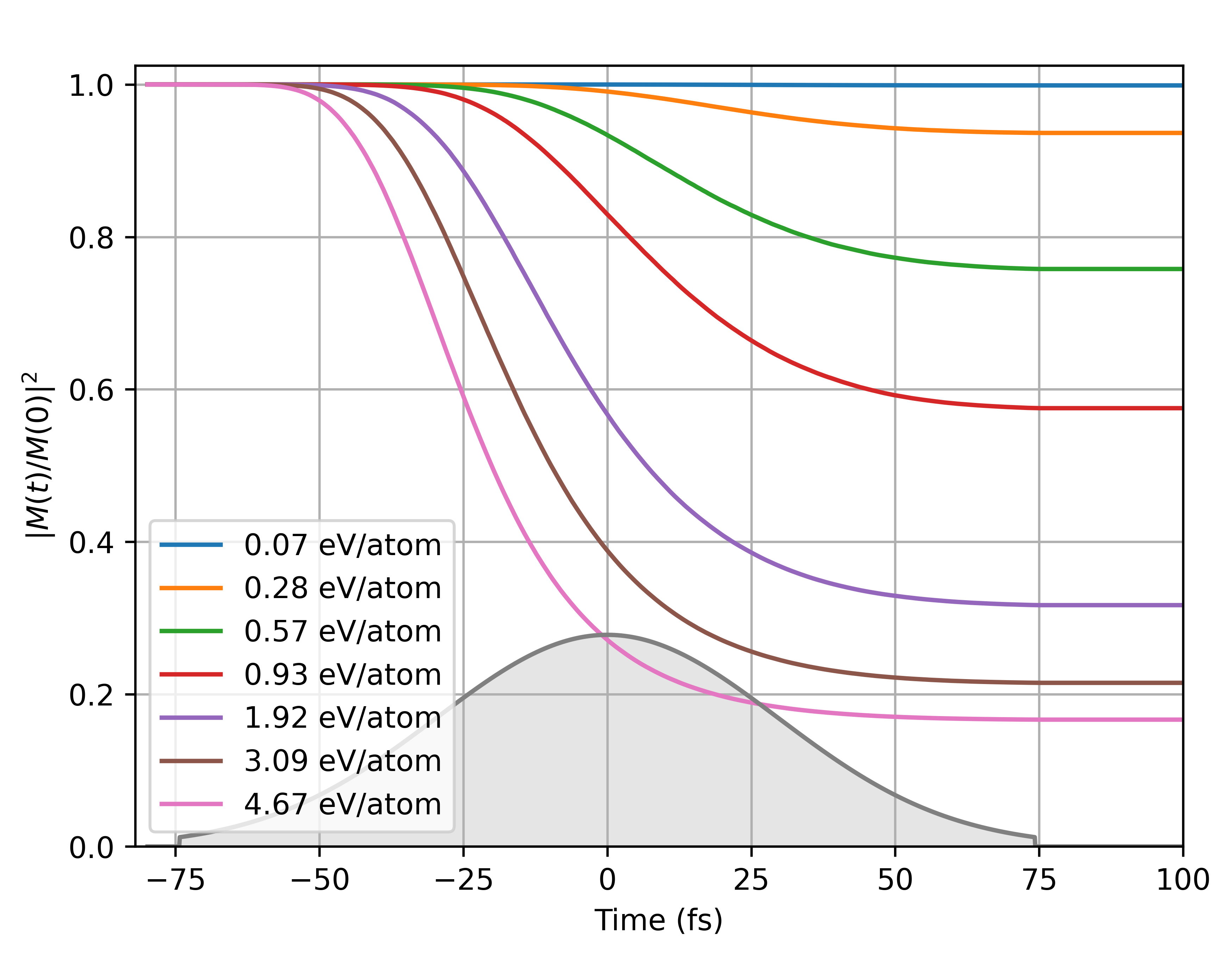}
\caption{\label{fig:MedgeSignal}%
Time dependence of squared normalized magnetization in bulk nickel obtained for the incoming photon energies around the $M_3$-edge case of Ni: 67 eV (left) and 68 eV (right). The curves obtained for different average absorbed doses are shown. The $\Delta$ was equal to 0.7 eV.}
\end{figure*}

\begin{figure*}
\includegraphics[width=\sizefour]{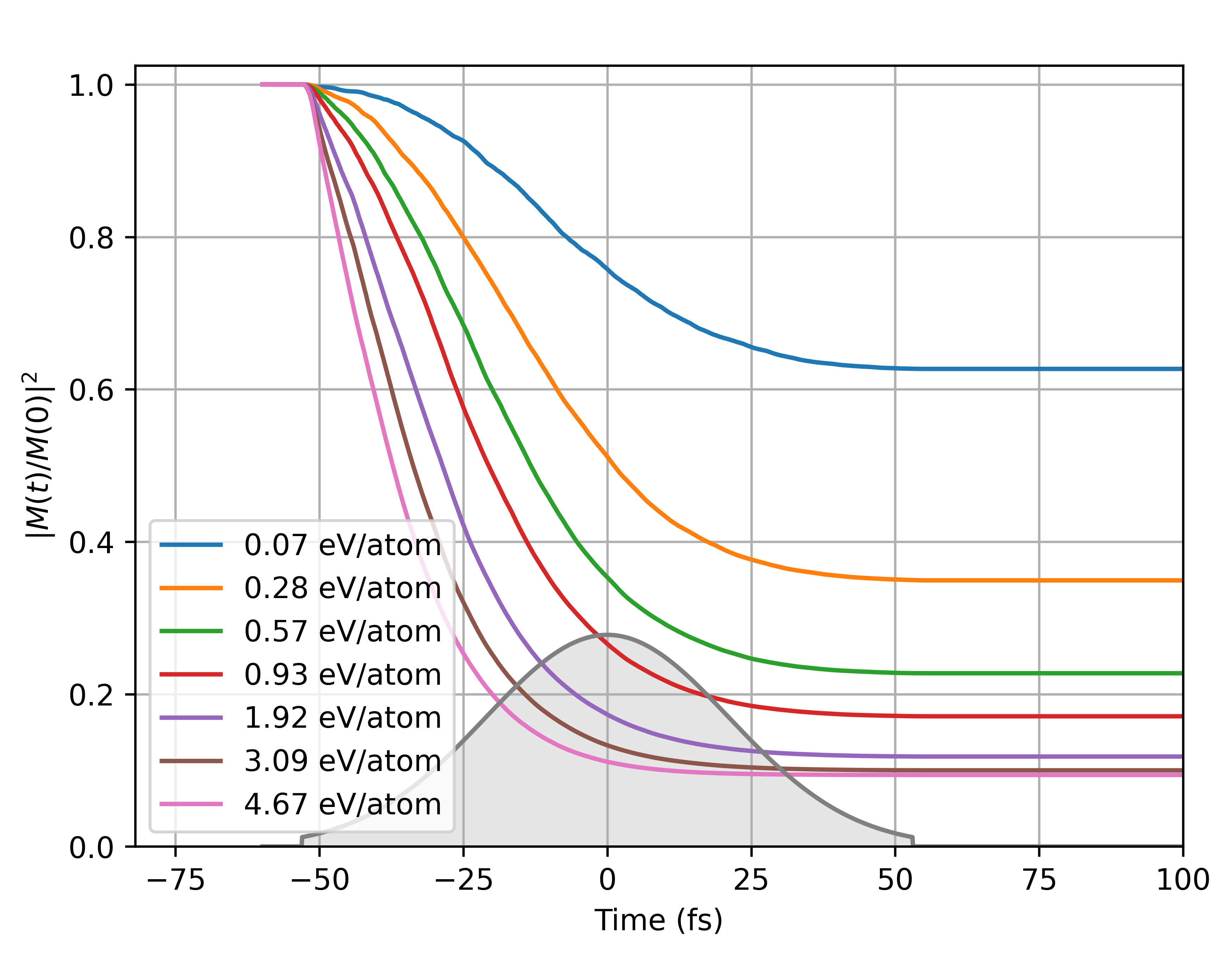}
\includegraphics[width=\sizefour]{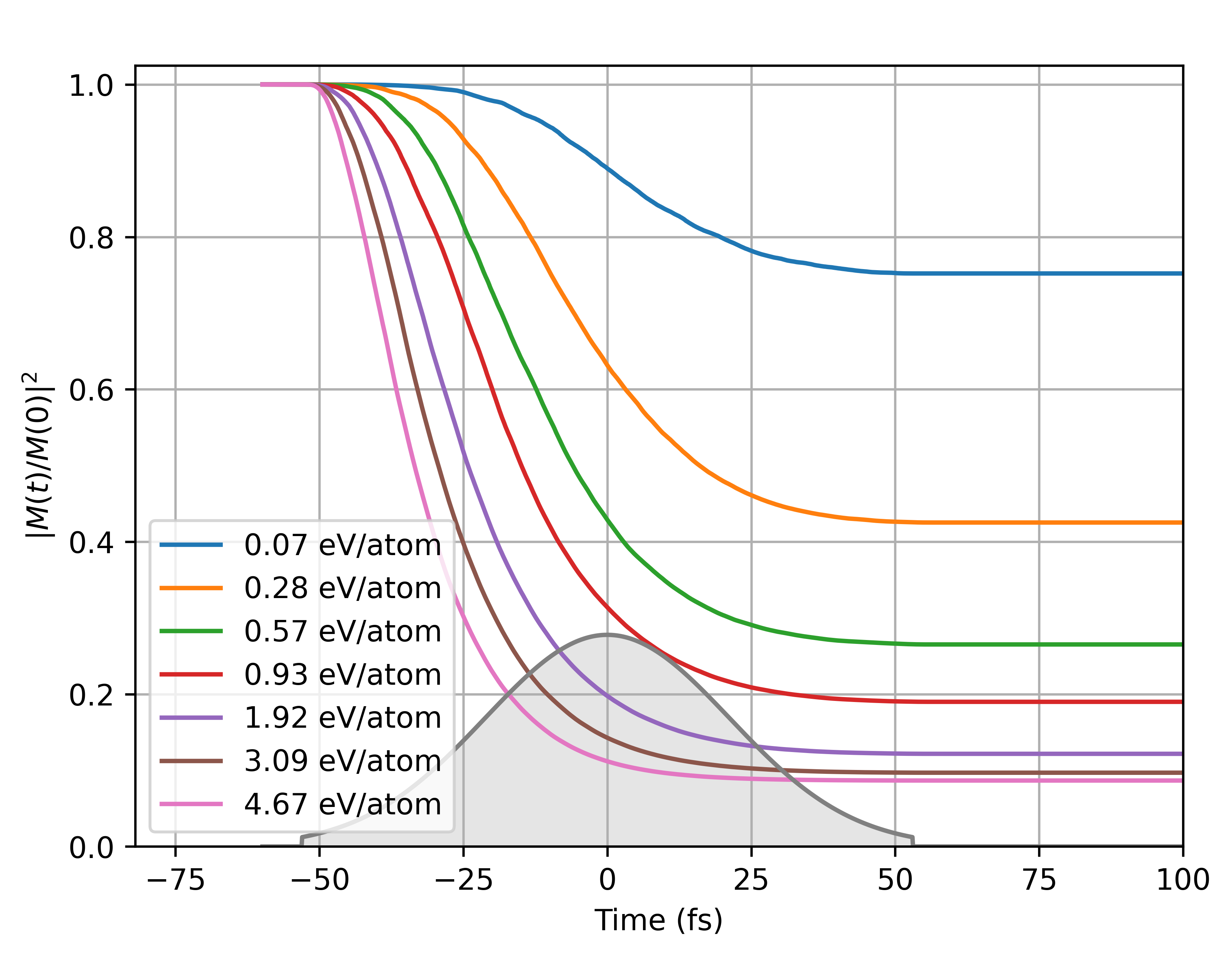}
\caption{\label{fig:LedgeSignal}%
Time dependence of squared normalized magnetization in bulk nickel obtained for the incoming photon energy around the $L_3$-edge case of Ni: 853 eV (left) and 854 eV (right). The curves obtained for different average absorbed doses are shown. The $\Delta$ was equal to 1.0 eV.}
\end{figure*}

Time evolution of squared generalized magnetization $M^2(t)$ (normalized to its initial value before the pulse at $t\rightarrow-\infty$; cf. also (\ref{eq:genmagnet})) for different absorbed doses is presented in Figures \ref{fig:MedgeSignal} and   \ref{fig:LedgeSignal}. The values of  $\Delta$ in Ni were taken from experimental measurements. They are:  $\Delta = 0.7$ eV for nickel $M$-edge \cite{OlsonSSC1980,MiedemaPCCP2019,ChangPRB2021} and $\Delta=1.0$ eV for nickel $L$-edge \cite{UchimotoJPS2001,CarvaEPL2009,UfuktepeXRS2011,GuDT2014}. Once can see that the decrease of magnetization becomes stronger with the increasing absorbed dose, and also strongly changes with incoming photon energy around the absorption edge. Interestingly, if the probed region in d-band includes the sharp peak in the DOS of spin-down electrons near the Fermi level (X-ray photon energies of 67 eV and 853 eV for M- and L-edge respectively; see Tab. \ref{tab:one}), the observed magnetization change is much stronger than in case when this peak is not included (X-ray photon energies of 68 eV and 854 eV for M- and L-edge respectively). The reason is that the peak ''provides'' a large number of unoccupied states for the resonant excitation from p-level, which leads to a stronger decrease of  the transient magnetization. Note that the decrease of magnetization is stronger for Ni than for Co (cf. Figure 4 from Ref. \cite{KapciaNPJ2022}  and Figure 3 from Ref. \cite{KapciaPRB2023} at the absorbed dose of 0.93 eV/atom). The reason is that cobalt DOS does not show such a peak close to the Fermi level, and the reduction of magnetization is, therefore, suppressed. This can also explain the lower Curie temperature for nickel than for cobalt.
%%%%%%%%
%%%%%%%
%%%%%%%%
%%%%%%%%
%%%%%%%%%%%%%%%%%%%%%%
\subsection{Calculation of the mSAXS signal}

\begin{figure}
\includegraphics[width=\sizetwo]{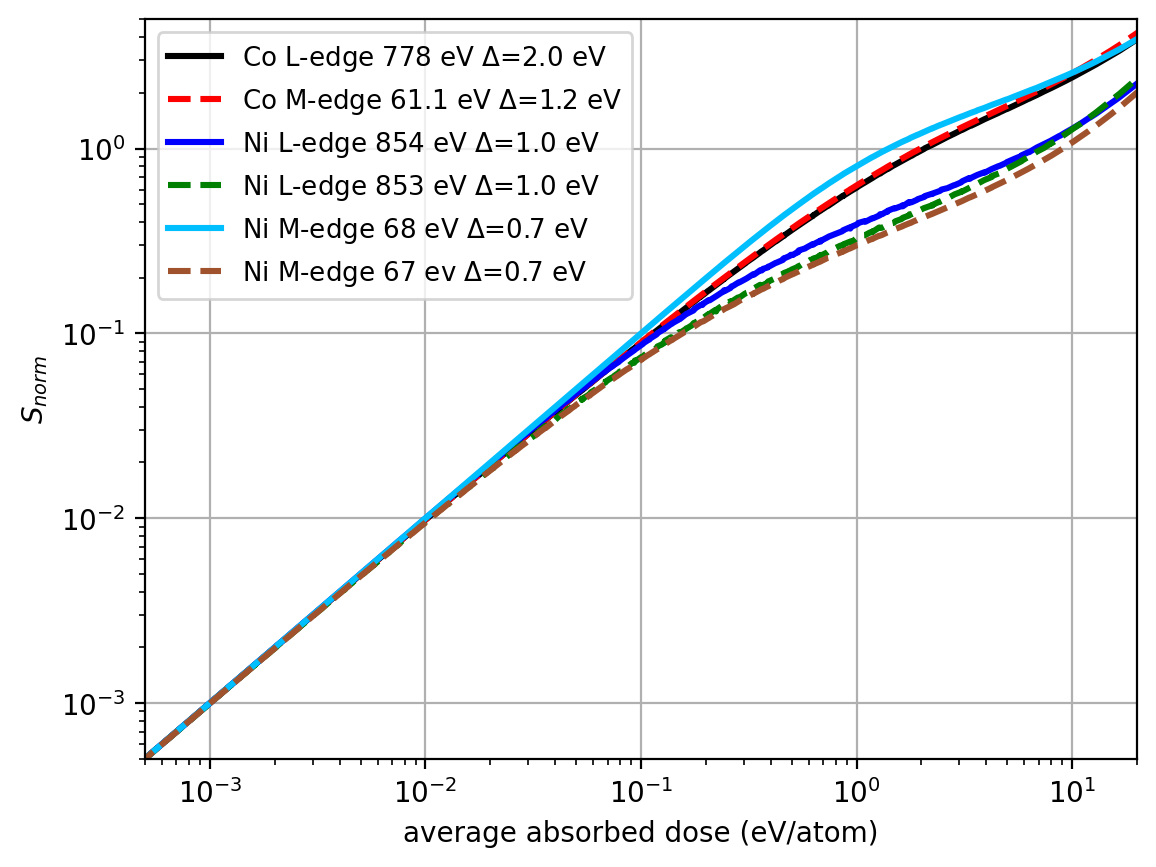}
\caption{\label{fig:mSAXS}%
Scattering efficiency $S_{\textrm{norm}}$ for nickel and for cobalt as a function of average absorbed dose for X-ray photon energies around  M and L absorption edges of Ni and Co.}
\end{figure}

Similarly as in \cite{KapciaNPJ2022}, we can calculate the mSAXS signal strength from the generalized magnetization. It is obtained as:
%%%%%%%%%%%%%%%%%
\begin{equation}\label{eq:signal}
S  = a \int M^2(t) I(t) dt, 
\end{equation}
%%%%%%%%%%%%%%%%%
where $I(t)$ is the X-ray pulse intensity and $a$ is a proportionality coefficient. Pulse fluence is then: $F =  \int I(t) dt$. It is proportional to absorbed dose, $D \propto F$, where the proportionality coefficient depends on the material parameters as well as on the photon energy. The dose dependence of the normalized signal strength, $S_{\textrm{norm}} = S(D)[D_0/S(D_0)]$ for the corresponding experimental $\Delta$ values is presented in Figure \ref{fig:mSAXS}. The normalization follows Ref. \cite{KapciaNPJ2022}, with the reference dose, $D_0=10^{-4}$ eV/atom for all considered cases. 

Similarly as observed for generalized magnetization, the signal strength strongly depends on the fact if the probed region in d-band includes or does not include the sharp peak in the DOS of spin-down electrons near the Fermi level - being distinctly higher in the latter case.  This explains also a stronger decrease of $S_{norm}$ for nickel than for cobalt. 
%%%%%%%
%%%%%%%
%%%%%%%%%%%
\section{Conclusions}

We provided theory predictions for electronic properties of X-ray irradiated Ni at photon energies close to $M_3$ or $L_3$ absorption edge, as well as for the resulting magnetization change and the mSAXS scattering strength. The results obtained indicate the same ultrafast demagnetization mechanism (caused by electronic excitation and relaxation) as in cobalt, occurring at a similar timescale. However, due to the difference in the DOS structure of the d-band, the degree of demagnetization for the equivalent dose would be higher in Ni than in Co. This finding is also consistent with the lower Curie temperature for nickel than for cobalt.

As in our previous studies on Co, we did not consider here atomic motion, and kept electronic band structure unchanged. This assumption does not hold for the case of high absorbed doses which may induce ultrafast  structural changes in irradiated materials. The model should then be developed further, enabling inclusion of atomic dynamics and of the transient band structure. 

Nevertheless, we expect that these theory predictions will inspire experimental studies on ultrafast X-ray induced demagnetization of nickel, a benchmark magnetic material of various applications.

%%%%%%%%%%%%%
\begin{acknowledgments}
The authors thank Leonard M\"uller and Andre Philippi-Kobs for helpful discussions at the early stages of the XSPIN model development. 
K.J.K. thanks the Polish National Agency for Academic Exchange for funding in the frame of the Bekker program (PPN/BEK/2020/1/00184).
V.T., A.L., S.M., B.Z. acknowledge the funding received from the Collaboration Grant of the European XFEL and the Institute of Nuclear Physics, Polish Academy of Sciences. 
\end{acknowledgments}
%%%%%%%%%%%%%

%\bibliographystyle{apsrev4-2}
\bibliography{biblio-nickel-arxiv}

\end{document}